\documentclass[aps,pre,reprint,10pt,superscriptaddress,showpacs]{revtex4-1}
\usepackage{amsfonts}
\usepackage{amsmath}
\usepackage{amssymb}
\usepackage{graphicx}
\usepackage{subfigure}
\usepackage{color}
\begin{document}
\title{Complex Korteweg-de Vries equation and Nonlinear dust-acoustic waves in  a magnetoplasma  with a pair of trapped  ions}
\author{A. P. Misra}
\email{apmisra@visva-bharati.ac.in; apmisra@gmail.com}
\affiliation{Department of Mathematics, Siksha Bhavana, Visva-Bharati University, Santiniketan-731 235, West Bengal, India}
\begin{abstract}
The nonlinear propagation of dust-acoustic (DA) waves in a magnetized dusty plasma with a pair of trapped ions is investigated. Starting from a set of hydrodynamic equations for massive dust fluids as well as kinetic Vlasov equations for ions,  and applying the reductive perturbation technique, a  Korteweg-de Vries (KdV)-like equation with a complex coefficient of nonlinearity is derived, which governs  the evolution of small-amplitude DA waves in plasmas. The complex coefficient arises due to   vortex-like distributions of both positive and negative ions.  An analytical as  well as numerical  solution of the KdV equation are obtained and analyzed with the effects of external magnetic field, the dust pressure as well as different mass and temperatures of positive and negative ions.
\end{abstract}
\date{5 January 2015}
\maketitle 
\section{Introduction}
Recently, there has been a renewed interest in investigating electrostatic disturbances in pair-plasmas  and, in particular,   plasmas with a pair of ions \cite{saleem2007,mahmood2009,misra2012,misra2013,eliasson2005,schamel2008,
schamel2005}. However, nonthermal pair plasmas may frequently occur not only in semiconductors in the form of electron and ion holes, but also in many astrophysical environments, e.g., pulsars, magnetars, as well as in the early universe, active galactic nuclei and supernova remnants in the form of electrons and positrons \cite{rees1983,burns1983,shukla1986,piran2004}. On the other hand, a number of experiments have been conducted to create pair-ion plasmas using fullerene as ion source  \cite{oohara2005}. Furthermore, it has been observed that the dust particles injected into a pair-ion plasma (e.g., $K^+/SF_6^{-}$ plasmas) can become positively charged when the number density of negative ions greatly exceeds that of electrons ($n_{n0}\gtrsim500n_{e0}$) \cite{kim2006,rosenberg2007}.  These pose  some possibilities  to investigate   collective behaviors as well as the formation of nonlinear coherent structures in  pair-ion plasmas  under controlled conditions. The formation of phase space holes in pure pair-ion plasmas \cite{eliasson2005} as well as ion holes in dusty pair-ion plasmas \cite{schamel2008} in the propagation of large amplitude electrostatic waves  have been investigated in the recent past in which  ions have been treated as trapped in self-created localized electrostatic potentials as prescribed by Schamel \cite{schamel1972}.  

In this paper we present a theoretical study on the formation and the dynamics of small-amplitude solitary structures in a dusty plasma composed of  charged dust particles and a pair  of   ions without electrons. In our theoretical model the massive charged dusts are described by a set of fluid equations, while the dynamics of both positive and negative ions are governed by kinetic Vlasov equations. Using the reductive perturbation technique we show that the evolution of small-amplitude electrostatic waves can be described by a Korteweg-de Vries (KdV)-like equation with a complex coefficient of the nonlinearity. A stationary  as well as numerical solutions of the KdV equation are obtained and analyzed with the effects of external magnetic field, the dust pressure as well as different mass and temperatures of ions.

\section{Basic Equations}
We consider the nonlinear propagation of dust-acoustic (DA)  solitary  waves in a magnetized dusty plasma which consists of positively or negatively charged mobile dusts and a pair of trapped ions with vortex-like distributions. The dust particles are assumed to have equal mass and constant charge. The collisions of all particles are also neglected compared to the dust plasma period. Furthermore, in dusty pair-ion plasmas the ratio of electric charge to mass of dust particles remains much smaller than those of positive and negative ions. We also assume that the size of the dust grains is small compared to the average interparticle distance.    The static magnetic field is considered along the $z$-axis, i.e., ${\bf B}=B_0\hat{z}$. While the dynamics of massive charged dusts in the propagation of DA waves ($v_{td}\ll v_p\ll v_{p,n}$, where $v_{tj}~(=\sqrt{k_BT_j/m_j})$ is the thermal velocity of  j-th species particles and $v_p$ is the phase velocity of the wave) is described by a set of fluid equations \eqref{cont-eq0} and \eqref{moment-eq0},  the dynamics of singly charged positive and negative ions are described by the Vlasov equations \eqref{vlasov-eq0}.
\begin{equation}
\frac{\partial n_d}{\partial t}+\nabla\cdot\left(n_d{\bf v}_d\right)=0, \label{cont-eq0}
\end{equation}
\begin{equation}
\frac{\partial {\bf v}_d}{\partial t}+\left({\bf v}_d\cdot\nabla\right){\bf v}_d=
\frac{q_d}{m_d}\left({\bf E}+{\bf v}_d\times{\bf B_0}\right)-\frac{\nabla P}{m_d n_d}, \label{moment-eq0}
\end{equation}
\begin{equation}
\frac{\partial f_j}{\partial t}+{\bf v}\cdot\nabla f_j-\frac{q_j}{m_j}\nabla\phi\cdot\frac{\partial f_j}{\partial {\bf v}}. \label{vlasov-eq0}
\end{equation}
The system of equations is then closed by the Poisson equation 
\begin{equation}
\nabla\cdot{\bf E}=4\pi e\left(n_p-n_n+\alpha Z_dn_d\right). \label{poisson-eq0}
\end{equation}
In Eqs. \eqref{cont-eq0}-\eqref{poisson-eq0}, $q_j,~n_j,~{\bf v}_j,~f_j$ and $m_j$   respectively, denote the charge, number density (with its equilibrium value $n_{j0}$), velocity, velocity distribution function,   and mass of $j$-species particles. Also, $q_d=\alpha z_de$ with $\alpha=\pm$ denoting for positively/negatively charged dusts and $z_d$   the charge state. Also,  ${\bf E}=-\nabla\phi$ is the electric field with $\phi$ denoting the electrostatic potential and $P$ is the dust thermal pressure given by the adiabatic equation of state $P/P_0=\left(n_d/n_{d0}\right)^{\gamma}$. Here, $\gamma=5/3$ is the adiabatic index for three-dimensional configuration and $P_0=n_{d0}k_BT_d$ is the equilibrium dust pressure with $k_B$ denoting the Boltzmann constant and $T_j$ the thermodynamic temperature of $j$-species particles. Furthermore, the ion densities are given by 
 \begin{equation}
 n_j=\int_{-\infty}^{\infty} f_j d{\bf v}. \label{density-eqn}
 \end{equation}
In what follows, we recast Eqs. \eqref{cont-eq0}-\eqref{poisson-eq0}   in terms of dimensionless variables. To this end the physical quantities are normalized according to $n_j\rightarrow n_j/n_{j0}$, $({\bf v},~{\bf v}_d)\rightarrow ({\bf v},~{\bf v}_d)/c_d$, $\phi\rightarrow e\phi/k_BT_p$ with $e$ denoting the elementary charge,    $f_j\rightarrow f_jv_{tp}/n_{j0}$, where $c_d=\sqrt {z_dk_BT_p/m_d}=\omega_{pd}\lambda_D$ is the dust-acoustic speed with $\omega_{pd}=\sqrt{4\pi n_{d0}z^2_d e^2/m_d}$ and $\lambda_D=\sqrt{k_BT_p/4\pi n_{d0}z_d e^2}$ denoting, respectively, the dust plasma frequency and the plasma Debye length.  The space and time variables are normalized by $\lambda_D$ and $\omega_{pd}^{-1}$ respectively. Thus, from Eqs.   \eqref{cont-eq0}-\eqref{density-eqn} we have following    set of normalized equations.  
\begin{equation}
\frac{\partial n_d}{\partial t}+\nabla\cdot\left(n_d{\bf v}_d\right)=0, \label{cont-eq}
\end{equation}
\begin{equation}
\frac{d {\bf v}_d}{d t}+\alpha\nabla\phi=\alpha\omega_c{\bf v}_d\times\hat{z}-\frac{5}{3}\sigma_dn_d^{-1/3}\nabla n_d, \label{moment-eq}
\end{equation}
 \begin{equation}
   \delta\frac{\partial f_j}{\partial t}+{\bf v}\cdot\nabla f_j-\zeta_j\frac{m_p}{m_j}\nabla \phi\cdot\frac{\partial f_j}{\partial {\bf v}}=0, \label{vlasov-eqn}
   \end{equation}
\begin{equation}
\nabla^2\phi=\mu_n n_n-\mu_p n_p-\alpha n_d, \label{poisson-eq}
\end{equation}
\begin{equation}
 n_j=\int_{-\infty}^{\infty} f_j d{\bf v}, \label{density-eqn1}
 \end{equation}
where $d/dt\equiv\partial/\partial t+{\bf v}_d\cdot\nabla$, $\alpha=\pm1$ for positively/negatively charged dusts,       $\omega_c=|q_d|B_0/m_d\omega_{pd}$ is the dust-cyclotron frequency normalized by the dust plasma frequency, $\sigma_d\equiv T_d/T_pz_d$, $\delta=\sqrt{z_dm_p/m_d}$, $\zeta_j=\pm1$ for positive/negative ions and $\mu_j=n_{j0}/Z_dn_{0}$ are the density ratios  $(j=p,n)$   which satisfy the following  charge neutrality condition at equilibrium:
\begin{equation}
\mu_p+\alpha=\mu_n. \label{charge-neutr}
\end{equation}  
\begin{figure*}[ht]
\centering
\includegraphics[height=3.0in,width=6.0in]{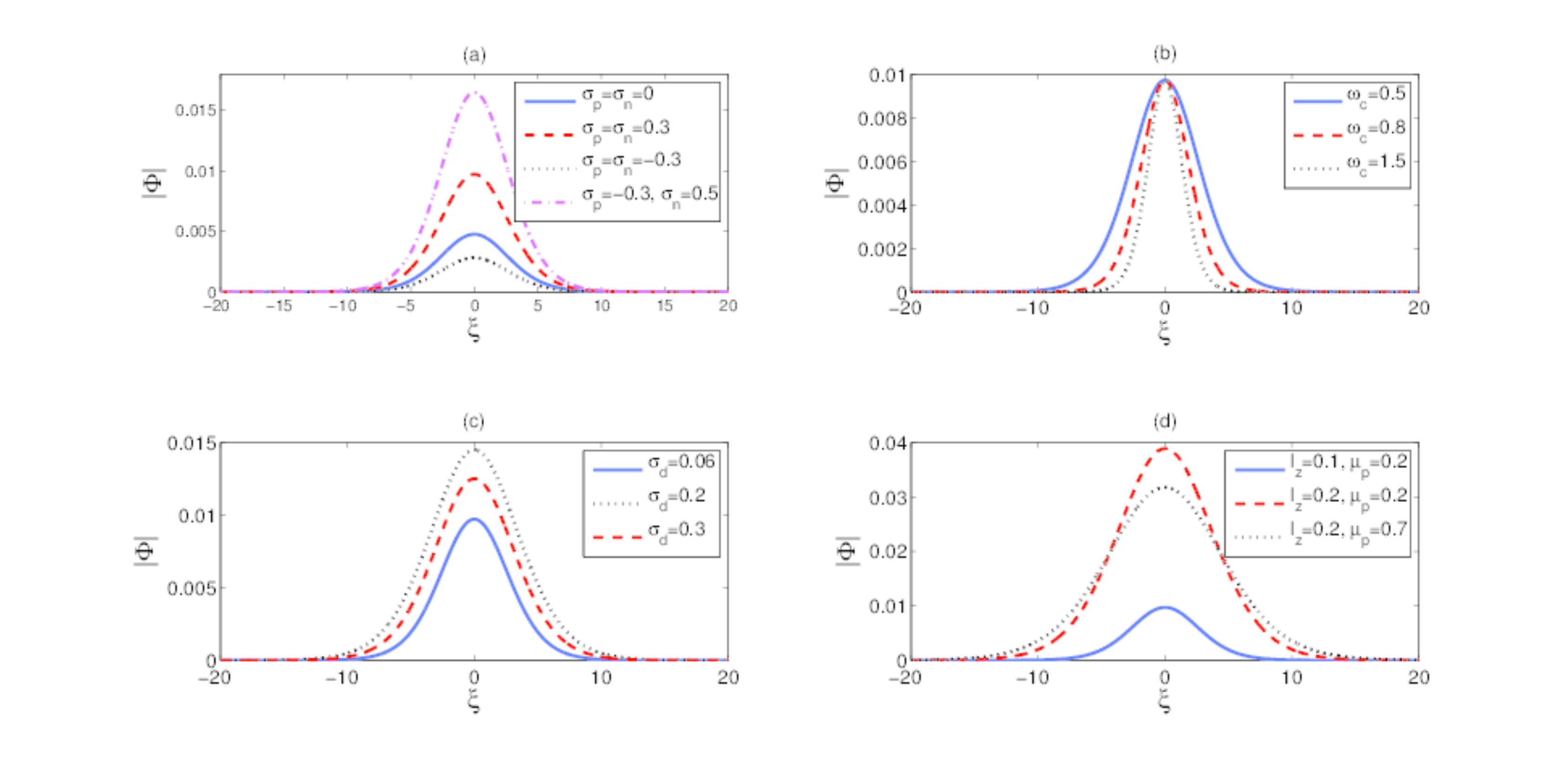}
\caption{Profiles of $|\Phi|$ given by Eq. \eqref{soliton-sol} are shown with respect to $\xi$ (and with $\sigma\sim1$) for different values of the plasma parameters as in the figure. The fixed parameter values for the subplots (a) to (d), respectively,    are  (a) $\alpha=1,~\mu_p=0.2,~l_z=0.1,~\omega_c=0.5,~\sigma_d=0.06$ and $u_0=0.1$, (b)    $\alpha=1,~\mu_p=0.2,~l_z=0.1,~\sigma_p=\sigma_n=0.3,~\sigma_d=0.06$ and $u_0=0.1$, (c) $\alpha=1,~\mu_p=0.2,~l_z=0.1,~\sigma_p=\sigma_n=0.3,~\omega_c=0.5,~\sigma_d=0.06$ and $u_0=0.1$, and (d) $\alpha=1,~\mu_p=0.2,~\sigma_p=\sigma_n=0.3,~\omega_c=0.5,~\sigma_d=0.06$ and $u_0=0.1$.}
\label{fig:figure1}
\end{figure*}
We neglect the ion inertial effects compared to the charged dusts, i.e., $\delta\rightarrow0$ in Eq. \eqref{vlasov-eqn}. The distribution functions $f_j$ for positive and negative ions, which are constant of motion of the Vlasov Eq. \eqref{vlasov-eqn}, are chosen \cite{schamel1972} for the excitation of localized solitary waves so that (i) they are continuous, and  both the free particle distributions are Maxwellian distribution  where $\phi\rightarrow0$ at $|\xi|\rightarrow\pm\infty$ and trapped particles are absent, (ii) both  trapped particle  distributions are  Maxwellian (with also negative temperatures). Thus, $f_j$ (for free and trapped particles) are  (with a suitable choice of the normalization constants) \cite{schamel1972,mamun1998a,mamun1998b,bandyopadhyay1999}
for positive ions
\begin{equation}
f_{pf}(v)=\frac{1}{\sqrt{2\pi}}\exp\left[-\frac{1}{2}\left(v^2+2\phi\right)\right],~|v|>\sqrt{-2\phi}, \label{dist-fpf}
\end{equation}
\begin{equation}
f_{pt}(v)=\frac{1}{\sqrt{2\pi}}\exp\left[-\frac{\sigma_p}{2}\left(v^2+2\phi\right)\right],~|v|\leq\sqrt{-2\phi}, \label{dist-fpt}
\end{equation}
and for negative ions
\begin{equation}
f_{nf}(v)=\sqrt{\frac{m\sigma}{2\pi}}\exp\left[-\frac{m\sigma}{2}\left(v^2-\frac{2\phi}{m}\right)\right],~|v|>\sqrt{2\phi/m}, \label{dist-fnf}
\end{equation}
\begin{equation}
f_{nt}(v)=\sqrt{\frac{m\sigma}{2\pi}}\exp\left[-\frac{1}{2}m\sigma\sigma_n\left(v^2-\frac{2\phi}{m}\right)\right],~|v|\leq\sqrt{2\phi/m}, \label{dist-fnt}
\end{equation}
where $m~(=m_n/m_p\gtrsim1)$ is the mass ratio, $\sigma~(=T_p/T_n\gtrsim1)$ is the temperature ratio and $\sigma_j$, for $j=p,n$,  measure the inverse of the trapped positive and negative ion temperatures which may be negative $(\sigma_j<0)$ corresponding to a depression in the trapped particle distribution. The case of $\sigma_j\rightarrow0$ represents the plateau (constant or flat-topped) and $\sigma_j\rightarrow1$ corresponds to the Boltzmann distribution of ions.
Next, integrating the particle distribution functions \eqref{dist-fpf}-\eqref{dist-fnt} over the velocity space, i.e., using Eq. \eqref{density-eqn1} we obtain the number densities $n_j$ for positive and negative ions as
\begin{equation}
n_p(\phi)=I(-\phi)+\frac{1}{\sqrt{|\sigma_p|}}\left\lbrace \begin{array}{llll}
e^{-\sigma_p\phi}~\text{erf}\left(\sqrt{-\sigma_p\phi}\right),& \sigma_p\geq0\\
\frac{2}{\sqrt{\pi}}W\left(\sqrt{\sigma_p\phi}\right),& \sigma_p<0,
\end{array}      \right. \label{np-phi}
  \end{equation}
\begin{equation}
n_n(\phi)=I(\sigma\phi)+\frac{1}{\sqrt{|\sigma_n|}}\left\lbrace \begin{array}{llll}
e^{\sigma\sigma_n\phi}~\text{erf}\left(\sqrt{\sigma\sigma_n\phi}\right),& \sigma_n\geq0\\
\frac{2}{\sqrt{\pi}}W\left(\sqrt{-\sigma\sigma_n\phi}\right),& \sigma_n<0,
\end{array}      \right. \label{nn-phi}
  \end{equation}
where $I(x)=\exp(x)\left[1-\text{erf}(\sqrt{x})\right]$. The error and Dawson functions    $\text{erf}(x)$  and $W(x)$ are, respectively,    given by
\begin{equation}
\text{erf}(x)=\frac{2}{\sqrt{\pi}}\int^x_0 e^{-t^2}dt,~~W(x)=e^{-x^2}\int^x_0 e^{t^2}dt. \label{integrals}
\end{equation}
In the small amplitude limit $\phi\ll1$, so that $\sigma\phi\ll1$,    we obtain from Eqs. \eqref{np-phi} and \eqref{nn-phi} the following expressions for the number densities \cite{schamel1972,mamun1998a,mamun1998b,bandyopadhyay1999}
\begin{equation}
n_p\approx1-\phi-\frac{4(1-\sigma_p)}{3\sqrt{\pi}}\left(-\phi\right)^{3/2}+\frac{1}{2}\phi^2, \label{ne}
\end{equation} 
\begin{equation}
n_n\approx1+\left(\sigma\phi\right)-\frac{4(1-\sigma_n)}{3\sqrt{\pi}}\left(\sigma\phi\right)^{3/2}+\frac{1}{2}\left(\sigma\phi\right)^{2}. \label{ni}
\end{equation} 
 
\begin{figure*}[ht]
\centering
\includegraphics[height=3.0in,width=6.0in]{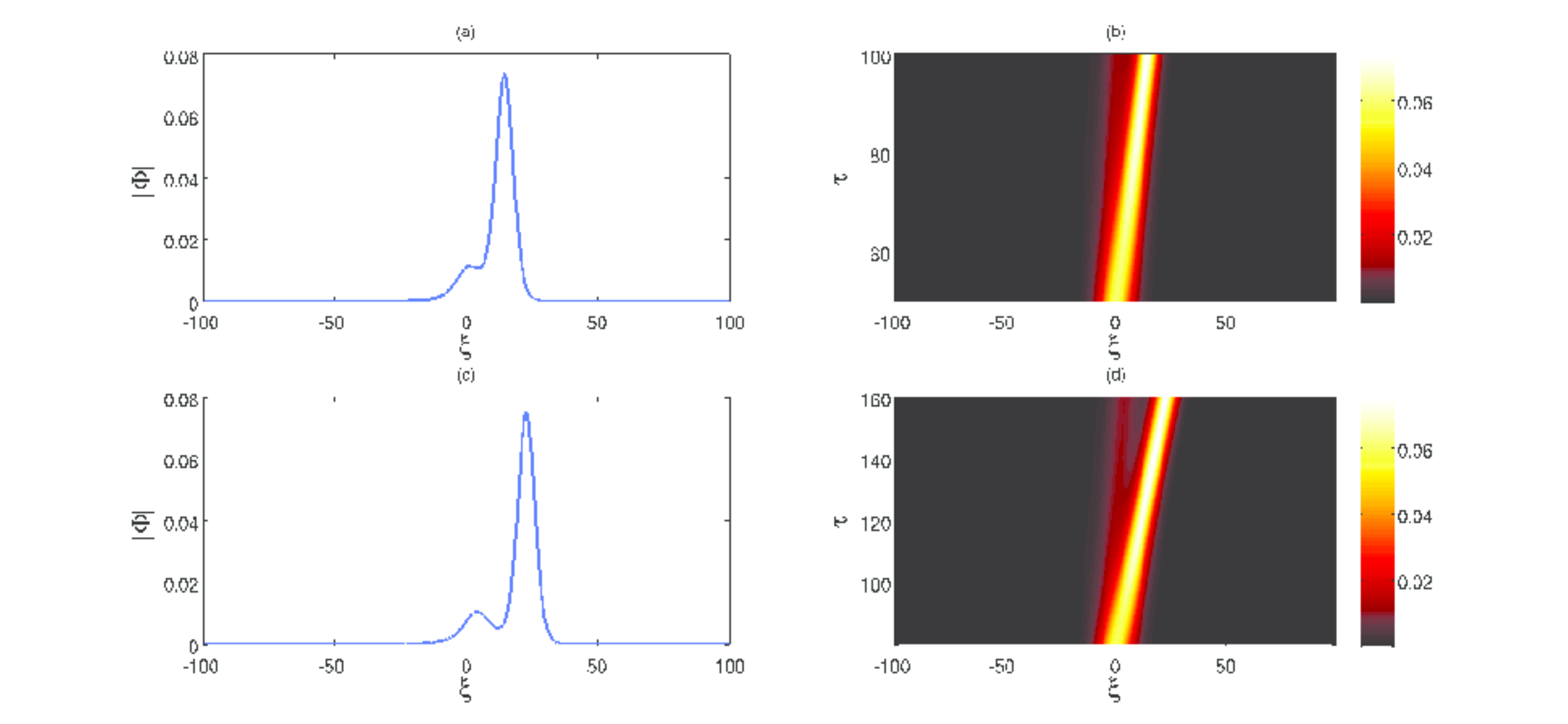}
\caption{ The space-time evolution of the soliton profile $|\Phi|$ [numerical solution of Eq. \eqref{kdv}] is shown at  $\tau=100$ [Subplots (a) and (b)] and $\tau=160$ [Subplots (c) and (d)]. While the plots (a) and (c) show the solton   profiles with space, plots   (b) and are the  corresponding contour plots.  The parameter values are the same as for the dashed line in Fig. \ref{fig:figure1}(d) and $\sigma\sim1$.     }
\label{fig:figure2}
\end{figure*}
\section{Evolution equation}
In order to derive the evolution equation for the DA waves, we transform the space and time variables  according to \cite{mamun1998a}
\begin{equation}
\xi=\epsilon^{1/4}\left(l_xx+l_yy+l_zz-Mt\right),~\tau=\epsilon^{3/4}t, \label{stretch-coord}
\end{equation}
where $\epsilon$ is a small parameter measuring the strength of nonlinearity. The dependent variables are expanded as \cite{mamun1998a}
\begin{equation}
\begin{split}
n=1+\epsilon n^{(1)}+\epsilon^{3/2} n^{(2)}+\cdots,\\
\phi=\epsilon \phi^{(1)}+\epsilon^{3/2} \phi^{(2)}+\cdots,\\
v_z=1+\epsilon v_z^{(1)}+\epsilon^{3/2} v_z^{(2)}+\cdots, \\
v_{x,y}=\epsilon^{5/4} v_{x,y}^{(1)}+\epsilon^{3/2} v_{x,y}^{(2)}+\cdots. \label{expansion}
\end{split}
\end{equation}
The anisotropy in Eq. \eqref{expansion} for the transverse velocity components of dust fluids  is introduced on the assumption that the dust gyromotion is a higher-order effect than the motion parallel to the magnetic field. 
Next, we substitute   Eqs.  \eqref{stretch-coord} and \eqref{expansion} into Eqs. \eqref{cont-eq}-\eqref{poisson-eq} and equate different powers of $\epsilon$ successively. In the lowest order $(\epsilon^{5/4})$, we obtain the following first-order quantities
\begin{equation}
n^{(1)}=\alpha\frac{l_z}{M}v_z^{(1)}=(\mu_p+\sigma\mu_n)\phi^{(1)}, \label{n1-vz1}
\end{equation}
\begin{equation}
v_{x,y}^{(1)}=\mp\frac{l_{y,x}}{\omega_c}\left(\frac{\partial\phi^{(1)}}{\partial\xi}+\frac{5}{3}\alpha\sigma_d\frac{\partial n^{(1)}}{\partial\xi}\right),
\end{equation}
and the dispersion relation for the nonlinear wave speed  given by
\begin{equation}
M=l_z\left(\frac{5}{3}\sigma_d+\frac{1}{\mu_p+\sigma\mu_n}\right)^{1/2}. \label{disp-rel}
\end{equation}
Replacing $M$ by $\omega/k$, one can obtain the same dispersion relation  after Fourier analyzing the linearized basic equations \eqref{cont-eq}-\eqref{poisson-eq}, i.e., assuming the perturbations as oscillations with the wave frequency $\omega$ and the wave number $k$.  We find that the phase speed $M$ (normalized by the DIA speed $c_d$) can be larger or smaller than the unity depending on the choice of the  parameter values. The value of $M$ increases with increasing values of both $l_z$ and $\sigma_d$. However, its values can slowly decrease with increasing values of the density ratios $\mu_j$ as well as the temperature ratio $\sigma$.    
From Eq. \eqref{cont-eq}, collecting the coefficients of     $\epsilon^{7/4}$ we obtain
\begin{equation}
M\frac{\partial n^{(2)}}{\partial\xi}=\frac{\partial n^{(1)}}{\partial\tau}+\sum_{j=x,y,z} l_j\frac{\partial v_j^{(2)}}{\partial\xi}. \label{cont-2nd}
\end{equation}
Similarly, equating the coefficients of    $\epsilon^{3/2}$ from the $x$- and $y$-components of Eq. \eqref{moment-eq}, and the coefficients of   $\epsilon^{7/4}$ from the $z$-component of Eq. \eqref{moment-eq} we successively obtain
\begin{equation}
\alpha v^{(2)}_{x,y}=\pm\frac{M}{\omega_c}\frac{\partial v_{y,x}^{(1)}}{\partial\xi}=\left[\frac{Ml_{x,y}}{\omega_c^2}+\frac{5}{3}\sigma_d\left(\mu_p+\sigma\mu_n\right)\frac{\partial^2\phi^{(1)}}{\partial\xi^2}\right], \label{moment-xy-2nd}
\end{equation}
\begin{equation}
M\frac{\partial v_z^{(2)}}{\partial\xi}=\frac{\partial v_z^{(1)}}{\partial\tau}+l_z\left(\alpha\frac{\partial \phi^{(2)}}{\partial\xi}+\frac{5}{3}\sigma_d\frac{\partial n^{(2)}}{\partial\xi}\right). \label{moment-z-2nd}
\end{equation}
From the coefficients of $\epsilon^{3/2}$ of Eq. \eqref{poisson-eq}, we obtain    an equation in which  $n^{(2)}$ is eliminated by the use of Eqs. \eqref{cont-2nd}, \eqref{moment-xy-2nd} and \eqref{moment-z-2nd}, and   the coefficient of $\phi^{(2)}$  vanishes by Eq. \eqref{disp-rel}. Thus, arranging the terms and using Eq. \eqref{n1-vz1} one obtains the following KdV-like equation 
\begin{equation}
\frac{\partial\Phi}{\partial\tau}+\left(A_p\sqrt{-\Phi}+A_n\sqrt{\Phi}\right)\frac{\partial\Phi}{\partial\xi}+B\frac{\partial^3\Phi}{\partial\xi^3}=0, \label{kdv0} 
\end{equation} 
where $\Phi\equiv\phi^{(1)}$. It follows that Eq. \eqref{kdv0} has a complex solution for $\Phi$. Typically, for $\Phi\sim r\exp(i\theta)$, where $r$  is a real function  of $\xi$ and $\tau$, and $\theta$ is a constant, one can have $\sqrt{-\Phi}=i\sqrt{\Phi}$. Thus, Eq. \eqref{kdv0}   can be written as  
\begin{equation}
\frac{\partial\Phi}{\partial\tau}+A\sqrt{\Phi}\frac{\partial\Phi}{\partial\xi}+B\frac{\partial^3\Phi}{\partial\xi^3}=0, \label{kdv}
\end{equation}
where  the coefficients of nonlinearity $(A\equiv A_n+iA_p)$ and dispersion $(B)$ are given by
\begin{equation}
A_j=\frac{\alpha}{\sqrt{\pi}}\frac{(1-\sigma_j)\mu_j}{M(\mu_p+\sigma\mu_n)}\left(\frac{T_p}{T_j}\right)^{3/2}, \label{nonlin-coef}
\end{equation}
\begin{equation}
B=\frac{l_z^2}{2M}\left[1+\frac{M^4(1-l_z^2)}{\omega_c^2l^4_z}\left(\mu_p+\sigma\mu_n\right)^2\right].
\end{equation}
The nonlinear coefficient $A$ becomes complex due to vortex-like distributions of two oppositely charged particles. In absence of one of them, $A$ becomes real, and one can then obtain solitary waves with positive or negative potential. A stationary soliton solution of Eq. \eqref{kdv} can easily be obtained with its absolute value as (For details see Appendix A)
\begin{equation}
|\Phi|=\Phi_0~\text{sech}^4\left[\left(\xi-u_0\tau\right)/W\right],\label{soliton-sol}
\end{equation}
where $u_0$ is a constant, and     $\Phi_0=\left(15u_0/8|A|\right)^2$  and $W=\sqrt{16B/u_0}$ are the amplitude and width of the soliton respectively.

\section{Results and Discussion}
We numerically analyze the solution \eqref{soliton-sol} with different plasma parameters as shown in Fig. \ref{fig:figure1}. Since $\sigma_j~(j=p,n)$ represents the reciprocal temperature  of the trapped positive and negative ions, and can be allowed from their negative to positive values  corresponding to different trapped particle  distributions, we consider negative, zero as well as positive values of $\sigma_j$.

 From Fig. \ref{fig:figure1}(a), it is seen that as $\sigma_j$ increases from $\sigma_j=0$ (corresponding to a constant or flat-topped distribution of ions)    to  $\sigma_j\sim1$  (corresponding to the Boltzmann distributions of ions), both  the amplitude and width of the soliton increase (See the solid and dashed lines).   Note here that the values of $\sigma_j>1$, for which the influence of the  trapped ions are inverted,  may be physically unrealistic as those correspond to a more steepened wave which can become unstable due to more peaked bump of the ion distributions. However, as the absolute value of $A$ starts increasing for $\sigma_j<0$, which corresponds to a depression in the trapped particle distribution, both the amplitude and width of the soliton are reduced (See the dotted line). The same can further be enhanced  for values of $\sigma_j$ satisfying $\sigma_p\sigma_n<0$ (See the dash-dotted line). 
 
Figure \ref{fig:figure1}(b) shows the soliton profile with the influence of the external magnetic field. Since $\omega_c$ contributes only to the dispersive coefficient $B$ of Eq. \eqref{kdv}, the effect of the magnetic field with increasing its intensity is to reduce   the width (without changing the amplitude) of the soliton. Thus, the external magnetic field makes the solitary structure more spiky. However,  for stronger magnetic fields with $\omega_c\gg1$, the width remains almost unaltered as in this case $B\sim l_z^2/2M$. 

The thermal effects of  charged dusts are shown in Fig. \ref{fig:figure1}(c). It is found that  the effect  of the dust thermal pressure $\sigma_d$  is to  enhance both   the  amplitude and width of the soliton.  The enhancement is due to the fact that as $\sigma_d$ increases, the values of $|A|$ $(B)$ decrease (increase), and hence the increase in both the amplitude and width. However, an opposite trend occurs by the effects of the positive to negative ion temperature ratio $\sigma$ (not shown in the figure). Typically, it reduces both the soliton amplitude and width significantly with a small increment of its value.

Figure \ref{fig:figure1}(d) exhibits the effects of the obliqueness of propagation $l_z$ and the relative (to dusts) concentration of  positive ions $\mu_p$. We find that both the amplitude and width of the soliton are greatly enhanced by a small increment of $l_z$ [Since  $A_j$  $(B)$ is inversely (directly) proportional to $l_z$]. However, as the positive ion concentration   increases, the amplitude gets reduced but the width increases.  

Next, we numerically solve Eq. \eqref{kdv} by the Runge-Kutta scheme with an initial condition of the form  $\Phi(\xi)=0.05~\text{sech}^4(\xi/10)\exp(-i\xi/15)$ and time step $d\tau=0.001$. The development of the wave form $|\Phi|$ after a finite interval of time   is shown in Fig. \ref{fig:figure2}. The parameter values are considered as the same as for the dashed line in Fig. \ref{fig:figure1}(d). It is  seen that the leading part of the initial wave steepens due to positive nonlinearity. As the time goes on the pulse separates into solitons and a residue due to the wave dispersion [See the subplots (a) and (b)]. It is found that  once the solitons are formed and separated, they propagate in the forward direction without changing their shape due to the nice balance of the nonlinearity and dispersion [See the subplots (c) and (d)].   

\section{Conclusion}
We have investigated the nonlinear propagation of dust-acoustic waves in a magnetized plasma which consists of warm positively charged dusts and  a pair of   free, as well as, trapped ions.  We have shown that the evolution of small-amplitude DA waves can be described  by a KdV-type equation with a complex coefficient of the nonlinearity. Such complex coefficient appears due to vortex-like distributions of both the ion species. The KdV equation is solved both analytically and numerically. The properties of the absolute value of $\Phi$ are only exhibited graphically. It is shown that while the external magnetic field only influences the width of the soliton, the trapped ion temperatures, the thermal pressures of ions and dusts, the relative concentration of positive ions   as well as the obliqueness of propagation have significant effects on both the amplitude and width of the solitons. We stress  that other solutions \cite{biswas2010,ganji2009,hizel2009,atif2010}  than those presented here of the complex KdV equation could of interest but beyond the scope of the present work. To conclude,  the present results  should be useful in understanding the nonlinear features of electrostatic localized disturbances in laboratory and space plasmas. 
 \section*{Acknowledgments} This work was partially supported by  the SAP-DRS (Phase-II), UGC, New Delhi, through sanction letter No. F.510/4/DRS/2009 (SAP-I) dated 13 Oct., 2009, and by the Visva-Bharati University, Santiniketan-731 235, through Memo No. REG/Notice/156 dated 07 January, 2014. APM thanks Dr. M. M. Panja of Department of Mathematics, Visva-Bharati, Santiniketan, India for some useful discussions. 
 
 \appendix 
\section{ Stationary solution of the KdV-like equation}
Equation \eqref{kdv}  is recast as
\begin{equation}
\frac{\partial\Phi}{\partial\tau}+A\sqrt{\Phi}\frac{\partial\Phi}{\partial\xi}+B\frac{\partial^{3}\Phi}{\partial\xi^{3}}=0. \label{eq0}
\end{equation}
Next, we apply the  transformation $\eta=\xi-u_0\tau$  to obtain from Eq. \eqref{eq0}
\begin{equation}
\frac{d}{d\eta}\left(B \ddot{\Phi}-u_0\Phi+\frac{2}{3}A\Phi^{3/2}\right)=0, \label{eq1}
\end{equation}
where the dot denotes differentiation with respect to $\eta$. Integrating Eq. \eqref{eq1} with respect to $\eta$ and using the boundary conditions $\Phi,~\ddot{\Phi}\rightarrow0$ as $\xi\rightarrow\pm\infty$ we get
\begin{equation}
B \ddot{\Phi}-u_0\Phi+\frac{2}{3}A\Phi^{3/2}=0. \label{eq2}
\end{equation}
Multiplying Eq. \eqref{eq2} by $2\dot{\Phi}$ and integrating once with respect to $\eta$, we obtain
 \begin{equation}
B \dot{\Phi}^2-u_0\Phi^2+\frac{8}{15}A\Phi^{5/2}=0, \label{eq3}
\end{equation}
where we have used the boundary conditions $\Phi,~\dot{\Phi}\rightarrow0$. From Eq. \eqref{eq3} we have
\begin{equation}
\dot{\Phi}=\pm\Phi\sqrt{\frac{u_0}{B}-\frac{8A}{15B}\sqrt{\Phi}},
\end{equation}
\begin{equation}
\text{or,~}\int\frac{d\Phi}{\Phi\sqrt{{u_0}/{B}-\left({8A}/{15B}\right)\sqrt{\Phi}}}=\pm \int d\eta,\label{eq4}
\end{equation}
which gives ($a=u_0/B$ and $b=8A/15B$)
\begin{equation}
 -\frac{4}{\sqrt{a}}\tanh^{-1}\sqrt{\frac{a-b\sqrt{\Phi}}{a}}=\pm\eta,
 \end{equation}
 \begin{equation}
 \text{or,~}\sqrt{\frac{a-b\sqrt{\Phi}}{a}}=\mp\tanh\left(\frac{\sqrt{a}}{4}\eta\right).\label{eq5}
\end{equation}
Thus, we obtain a soliton solution  of Eq. \eqref{kdv} as
\begin{equation}
1-\tanh^2\left(\frac{\sqrt{a}}{4}\eta\right)= \frac{b}{a}\sqrt{\Phi},
\end{equation}
\begin{equation}
\text{or,~}\Phi=\left(\frac{15u_0}{8A}\right)^2\text{sech}^4\left(\sqrt{\frac{u_0}{16B}}\eta\right).\label{eq6}
\end{equation}



\begin{thebibliography}{50}
\bibitem{saleem2007} H. Saleem, A criterion for pure pair-ion plasmas and the role of quasineutrality in nonlinear dynamics,  Phys. Plasmas 14 (2007) 014505.
\bibitem{mahmood2009} S. Mahmood, H.U. Rehman,  H. Saleem, Electrostatic Korteweg–de Vries solitons in pure pair-ion and pair-ion–electron plasmas,  Phys. Scr. {80} (2009) 035502.
\bibitem{misra2012} A.P. Misra, N.C. Adhikary,   P.K. Shukla, Ion-acoustic solitary waves and shocks in a collisional dusty negative-ion plasma,  Phys. Rev. E {86} (2012)   056406.
\bibitem{misra2013} A.P. Misra,  N.C. Adhikary, Electrostatic solitary waves in dusty pair-ion plasmas,  Phys. Plasmas {20} (2013) 102309.
\bibitem{eliasson2005} B. Eliasson,   P.K. Shukla, Solitary phase-space holes in pair plasmas,  Phys. Rev. E 71 (2005)  046402.
\bibitem{schamel2008} H. Schamel, Ion holes in dusty pair plasmas,  J. Plasma Phys. 74  (2008) 725-731.
\bibitem{schamel2005} H. Schamel, L. Luque, Kinetic theory of periodic hole and double layer equilibria in pair plasmas,   New  J. Phys. 7 (2005) 69.
\bibitem{rees1983} M.J. Rees,  The Very Early Universe, edited by G.B. Gibbons, S.W. Hawking,  S. Sikias, Cambridge University Press, Cambridge, 1983.
\bibitem{burns1983} Positron-Electron Pairs in Astrophysics, edited by M.L. Burns
{\it et al.}, AIP, New York, 1983.
\bibitem{shukla1986}  P.K. Shukla, N.N. Rao, M.Y. Yu, N.L. Tsintsadze, Relativistic nonlinear effects in plasmas, Phys. Rep. 138   (1986) 1-149.
\bibitem{piran2004} T. Piran, The physics of gamma-ray bursts,  Rev. Mod. Phys. 76 (2004) 1143.
\bibitem{oohara2005} W. Oohara, D. Date,  R. Hatakeyama, Electrostatic waves in a paired fullerene-ion plasma,  Phys. Rev. Lett. {95} (2005) 175003.
\bibitem{kim2006} S-H. Kim, R.L. Merlino, Charging of dust grains in a plasma with negative ions,  Phys. Plasmas  {13} (2006) 052118. 
\bibitem{rosenberg2007} M. Rosenberg,  R.L. Merlino, Ion-acoustic instability in a dusty negative ion plasma, Planet. Space Sci. 55 (2007) 1464-1469.
\bibitem{schamel1972} H. Schamel, Stationary solitary, snoidal and sinusoidal ion acoustic waves,    Plasma Phys. 14 (1972) 905-924.
\bibitem{mamun1998a} A.A. Mamun, Nonlinear propagation of ion-acoustic waves in a hot magnetized plasma with vortexlike electron distribution,  Phys. Plasmas 5 (1998) 322.
\bibitem{mamun1998b} A.A. Mamun, Solitary Waves in a Three-component Dusty Plasma with
Trapped Ions,  Phys. Scr. 57 (1998) 258-260.
\bibitem{bandyopadhyay1999} A. Bandyopadhyay, K.P. Das, Stability of solitary waves in a magnetized non-thermal plasma with warm ions,  J. Plasma Phys. 62 (1999) 255-267.
\bibitem{biswas2010} A. Biswas, E. Zerrad, A. Ranasinghe, Dynamics of solitons in plasmas for the complex KdV equation with power law nonlinearity, Appl. Math.   Comput. 217 (2010) 1491–1496.
\bibitem{ganji2009} Z.Z. Ganji, D.D. Ganji, H. Bararnia, Approximate general and explicit solutions of nonlinear BBMB equations by Exp-Function method, Appl. Math. Modelling 33 (4) (2009) 1836–1841.
\bibitem{hizel2009} E. Hizel, Group invariant solutions of complex modified Korteweg–de Vries equation, Int.   Math.  Forum 4 (2009) 1383–1388.
\bibitem{atif2010} S. Atif, D. Milovic, A. Biswas, 1-Soliton solution of the complex KdV equation in plasmas with power law nonlinearity and time-dependent coefficients, Appl. Math.   Comput. 217 (2010) 1785–1789.




%
%
%
%











\end{thebibliography}
\end{document}